\documentclass[letterpaper]{article}

\usepackage[T1]{fontenc}

\usepackage{geometry}
\usepackage{setspace}

\usepackage{achemso}
\setkeys{acs}{articletitle = true}

\usepackage{graphicx}
\usepackage{float}
\newfloat{scheme}{htbp}{los}
\floatname{scheme}{Scheme}
\floatname{chart}{Chart}
\newfloat{graph}{htbp}{loh}
\usepackage[numbers]{natbib}
\usepackage{graphicx}
\usepackage{dcolumn}
\usepackage{bm}
\usepackage{amsmath}
\usepackage{amssymb}
\usepackage{gensymb}
\usepackage{siunitx}
\usepackage{xcolor}

\usepackage{chemformula} 
\usepackage[version = 4]{mhchem} 

\usepackage{authblk}
\author[1]{Artur~Avdizhiyan}
\affil[1]{Institute of Fundamental and Frontier Sciences, University of Electronic Science and Technology of China, Chengdu 610054, P. R. China}
\author[2]{Ilya~Razdolski*}
\affil[2]{Photonics Research Centre, Universiti Malaya, 50603 Kuala Lumpur, Malaysia}
\author[3]{Anton~Yu.~Bykov*}
\affil[3]{Nanophotonics Centre, Cavendish Laboratory, Department of Physics, University of Cambridge, CB3 0US, UK}

\title{Non-equilibrium quantum plasmonics in nanoparticle-on-mirror nanocavities}
\date{*Email: ilya@um.edu.my, ab3228@cam.ac.uk}

\begin{document}

\maketitle

\begin{abstract}
    We develop a novel approach to ultrafast optical modulation of quantum-mechanical phenomena at the interface of plasmonic metals. Focusing on efficient and versatile nanoparticle-on-mirror plasmonic nanocavities, we discuss indirect control of plasmonic properties through laser-induced ballistic hot electron injection. Overcoming the limitations precluding the observations of laser-driven mesoscopic phenomena in the time domain with state-of-the-art amplified sources, our proposed experimental approach can be readily realized without irreversible optical damage and holds immense potential for the future development of ultrafast electrodynamics in nanogaps, applications in photochemistry and nanoscale control of quantum emitters. Agreeing with the results of numerical simulations, an intuitive microscopic model for the proposed time-dependent mesoscopic electrodynamics facilitates the analysis of the temperature-induced modulation of quantum plasmonic properties in a broad parameter space. Our work expands the realm of quantum nanophotonics onto non-equilibrium electronic systems and facilitates the development of ultrafast methods in active plasmonics.
\end{abstract}

\section*{Keywords}

Quantum plasmonics, Hot electron dynamics, Nanoparticle-on-Mirror (NPoM), Feibelman parameters, Ultrafast Electrodynamics



The recent development of quantum plasmonics, an intriguing branch of nanophotonics 
at nano- and mesoscale, has led to multiple investigations of nanoobjects \cite{TameNatPhys2013,MarinicaSci2015,FitzgeraldIEEE2016,ChristensenPRL2017,DingJPCM2018,https://doi.org/10.1002/apxr.202400144}. Relevant physical processes and their spatial scales were identified, indicating different regimes where nonclassical phenomena govern the optical response of highly confined plasmonic modes\cite{Savage2012, scholl2013}. Lately, 
a physically rich landscape of quantum nanoplasmonics was complemented with the analysis of surface geometry and crystalline structures \cite{RodriguezEcharri2021}. 

The optical response of mesoscopic quantum systems is often quantified with the Feibelman parameters\cite{Feibelman76,Feibelman82}. Providing a convenient formalism for the plasmon dispersion and a clear physical meaning, they are routinely used for the characterization of quantum-induced optical non-locality. 
Despite numerous computational methods, little is known about manipulating the quantum properties of plasmonic nanostructures through active control in the time-domain, and experimental spectroscopic evidence still remains scarce. This underscores a significant gap to the conventional, local plasmonics employing the laser-driven tuning of plasmon resonances through hot carrier excitation \cite{MacDonaldNatPhot2009,MetzgerACSPhot2016,KimNanop2022}. Although laser-induced electron dynamics in metals is extensively studied, the application of hot carrier concepts to quantum phenomena at the nanoscale remains elusive due to the weakness of these perturbations to the non-local response, and the technical difficulties in avoiding optical damage. 

In this Letter, we argue that 
the observation of non-local effects in dynamics is feasible within the currently existing experimental approaches in ultrafast active plasmonics. We develop a microscopic theory relating the quantum-driven optical non-locality in plasmonic systems to the electron temperature and analyze the impact of temperature and Feibelman non-locality on equal footing. Numerical simulations demonstrate the feasibility of laser-induced modulation of the Feibelman parameters in non-equilibrium plasmonic systems and provide quantitative estimates of the hot electron-driven optical non-locality. We propose an experimental concept that overcomes the limitations of direct methods and enables the observation and control of mesoscopic quantum phenomena, paving the way towards active quantum plasmonics.

We introduce the optical non-locality through the concept of the surface-response function \cite{PhysRevLett.115.137403, Yang2019General}. In real metals, the free carrier density is not localized within the sharp boundary of ionic background but is allowed to "spill out" \cite{PhysRevB.7.3541}, changing the surface electrostatic barrier. When subjected to high-frequency electromagnetic radiation, this electron "spill-out" modifies the dynamic screening at the interface and thus the optical response \cite{PhysRevB.36.7378}. This behavior can be self-consistently treated using time-dependent density functional theory (TD-DFT), or through the introduction of the Feibelman parameters $d_{\perp}$, $d_{\|}$ which parametrize the first moments of the induced charge and current density\cite{Feibelman82}. 
Multiple physical processes, including losses due to Landau damping\cite{Feibelman82,RodriguezEcharri2021,Babaze2023} are encased in $\operatorname{Im} d_{\perp}$ which is thus paramount for surface chemistry \cite{Khurgin2024elight} and development of high-performance plasmonic devices, as it reflects fundamental limits of field enhancement in plasmonic antennae \cite{Khurgin2017}.

While full quantum treatment of optical non-locality in the excited state represents a significant challenge, we outline possible mechanisms that could govern the predicted effects and estimate their strength by adapting the state-of-the-art TD-DFT approach\cite{PhysRevB.52.14219}. Occupying the infinite half-space ($z<0$), a prototypical plasmonic metal (gold) is described in the local density approximation (LDA) by the bulk density corresponding to $r_s = 3.0$. The external potential in the nonretarded case is $\psi_{ext} = -2\pi z$. The self-consistent solution for the induced electron density $\delta n$ at a given frequency $\omega$ is: 

\begin{equation}
\begin{split}
\delta n = \int dz' \chi(z, z', \omega)[\psi_{ext}(z', \omega) \\+ \psi_{ind}(z', \omega) + \delta V_{xc}(z', \omega) ],
\end{split}
\end{equation}
where $\chi(z, z^{\prime}, \omega)$ is a jellium first-order response function in the long wavelength limit, $\psi_{ind}$ is the induced Coulomb potential, and $\delta V_{xc}$ is the induced exchange-correlation contribution prescribed in adiabatic TD-LDA. Importantly, following Ref.~\cite{PhysRevB.52.14219} we incorporate additional screening imposed by the occupied d-band of gold, as a semi-infinite slab with dielectric function $\varepsilon_d(\omega)$, occupying $z < 0$ space, so that the induced Coulomb potential satisfies:

\begin{equation}
\left(\varepsilon_d(z, \omega)\frac{d\psi_{ind}}{dz}\right)' = -4\pi\,\delta n
\label{epsilon}
\end{equation}

However, instead of using empirical values\cite{PhysRevB.52.14219, Yang2019General,RodriguezEcharri2021} of $\varepsilon_d(\omega)$ we utilise a calculated state-of-the-art temperature-dependent expression previously used in time-domain studies \cite{Karaman2024NCom, PhysRevB.85.235403}. 
Additionally, we ensure consistent treatment of the electron interaction within the groundstate and response calculations by introducing static (yet temperature-dependent) screening $\varepsilon$ obtained from the same model in the low-frequency limit into the self-consistent solution of the Poission equation for the groundstate density.

Pure LDA calculations fail to correctly replicate the surface properties of gold, resulting in work function $W$ more than an electron-volt lower than the experimental values\cite{Yang2019General, RodriguezEcharri2021} and prompting the exploration of phenomenological artificial potentials instead \cite{RodriguezEcharri2021}.
Nevertheless, incorporating static screening and "stabilisation" of the jellium model \cite{PhysRevB.42.11627} we obtain a surprisingly good agreement of $W=5.37$~eV with the experiment ($5.3-5.5$ eV)\cite{SACHTLER1966221}. The introduction of static screening pushes the free electron density into the interior of the metal. This "spill-in" behaviour, expected for the noble metals \cite{PhysRevLett.115.137403}, is quantified by the Feibelman parameters\cite{Feibelman82} introduced as:

\begin{equation}
    \left. d_{\perp}=\int z\frac{d}{dz}E_z(z,\omega)~dz  \middle/ \int \frac{d}{dz}E_z(z,\omega)~dz \right.
\end{equation}

In practice, we compute  $d_{\perp}$ from the induced free electron density using the dynamic force sum rule\cite{PhysRevB.52.14219} (Supporting Information S1.2). 
We note that this process can be sidestepped by utilising the notion of nonlocal reflection coefficient \cite{PhysRevB.52.14219, RodriguezEcharri2021}. The latter, however, requires computations at finite momenta and fitting \cite{RodriguezEcharri2021} which we avoid in favor of simplicity and computational stability. 
Varying the d-band screening as a function of temperature in the ground state and response calculations we also modify Drude damping phenomenologically by computing the response kernels at complex frequencies $\omega \pm 0.5i\gamma(T_e)$ \cite{PhysRevB.52.14219}. 

To assess the role of the 
electronic temperature $T_e$ in both bulk and nonlocal surface response, we implement the state-of-the-art two-band model of the dielectric function of gold\cite{PhysRevB.12.557, PhysRevB.85.235403, Karaman2024NCom}. It independently incorporates the change in Drude damping due to electron temperature-dependent fractional Umklapp electron scattering \cite{PhysRevB.16.5277} and lattice temperature-dependent electron-phonon scattering, as well as broadening of the interband transitions\cite{doi:10.1021/acs.nanolett.3c00063, BykovRothSartorelloSalmonGamboaZayats+2021+2929+2938}. 
The total dielectric function consists of an interband contribution accounting for the transitions around the $X$ and $L$ high symmetry points as well as higher-order transitions prescribed by $\varepsilon_{\infty}$, and an intraband Drude term ($\varepsilon_s$), so that
$\varepsilon(\omega) = \varepsilon_s(\omega) + \varepsilon_d(\omega) - 1$:

\begin{align} \label{eq:rosei:eps}
    \operatorname{Im}[\varepsilon_{d}(\omega, T_e)] = &(\hbar\omega)^{-2}[A_X J_{X}(\omega, T_e)+ \\ A_{L_4^+} &J_{L_4^+}(\omega, T_e)+A_{L_{5+6}^+} J_{L^+_{5+6}}(\omega, T_e)], \notag \\
    \varepsilon(\omega) = \varepsilon_{\infty} + &\varepsilon_d(\omega, T_e)-\frac{\omega_p^2}{\omega^2+i\omega\gamma(\omega,T_e)}. 
\end{align}

Here $\omega_p = 9.02$~eV is the plasma frequency, $A_i$ are the spectral weights of different interband transitions, and $J_{i}$ are the joint densities of states calculated by integrating the energy distribution of the joint density of states (EDJDOS) with the Fermi-Dirac occupation factors at a given $T_e$ \cite{PhysRevB.85.235403, PhysRevB.12.557}. The damping $\gamma(\hbar\omega,T_e) = \gamma_{e-e}(\hbar\omega,T_e)+\gamma_{e-ph}$ contains the fractional Umklapp electron-electron and electron-phonon scattering terms. The former is given by:

\begin{equation}\label{eq:rosei:gamma}
    \gamma_{e-e}(\hbar\omega,T_e) = \frac{\pi^3\beta\Delta}{12\hbar E_F}\left[(k_BT_e)^2+(\hbar\omega/2\pi)^2\right]
\end{equation}
where $\Delta = 0.75$ and $\beta = 0.55$ are dimensionless constants characterising electron-electron scattering in noble metals \cite{PhysRevB.16.5277} and $E_F = 5.53$~eV is the Fermi energy. The band parameters are taken from Ref.~\cite{PhysRevB.85.235403}, and $A_i$ are determined by fitting\cite{Karaman2024NCom, PhysRevB.85.235403} to the room temperature dielectric function of gold \cite{JohnsonChristy1972}(See Supporting Information S1.2, Fig.~S6).
To match the reported room temperature static limit for d-electron screening \cite{PhysRevB.69.195416}, we chose $\varepsilon_{\infty} \approx 2.8$, similar to previous works \cite{JohnsonChristy1972, Neira2015}.

\begin{figure}
  \centering
  \includegraphics[width=\linewidth]{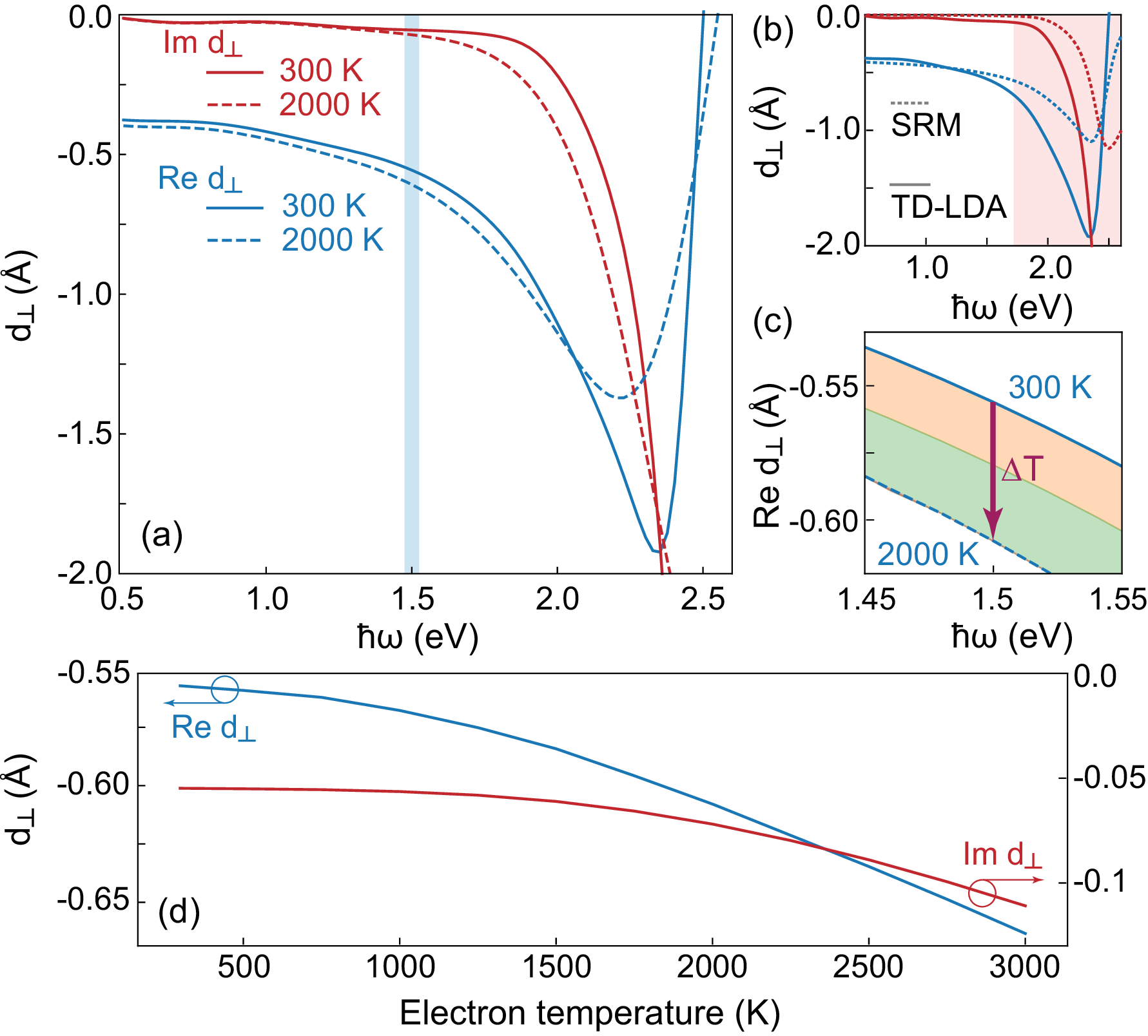}
    \caption{
    Electron temperature-dependent nonlocal surface response. 
    (a) Computed Feibelman parameter $d_{\perp}$ at a gold-vacuum interface at 300~K and 2000~K. 
    (b) Comparison of TD-LDA results (solid) with the specular reflection model (SRM, dotted). The red-shaded area indicates the region where interband transitions dominate. 
    (c) Zoomed-in view of (a, blue-shaded area) around $\hbar\omega=1.5$~eV. 
    The shaded areas in (c) indicate contributions from static (orange) and dynamic (green) screening to the total change of $d_{\perp}(\omega, T_e)$ upon heating from 300~K (solid line) to 2000~K (dashed line). The damping contribution (red) is invisible at the current scale.
    (d) Dependence of d$_{\perp}$($\hbar\omega=1.5$~eV) on the electron temperature.
    }
    \label{fig:theory}
\end{figure}

We calculate $d_{\perp}$ at the gold-vacuum interface (Fig.~\ref{fig:theory}a) at $T_e=300$~K and $T_e=2000$~K. The negative sign of $\operatorname{Re}~d_{\perp}$ at low frequencies signifies the "spill-in" electronic behaviour typical for noble metals. Qualitatively replicating the expected behaviour, the room temperature results (solid line) show a pronounced dip in $\operatorname{Re}~d_{\perp}$ around the surface plasmon frequency (2.5~eV). At elevated electron temperatures (dashed line), the dip is smeared out, reducing $|\operatorname{Re}~d_{\perp}|$ near the interband transitions. An extension of these calculations to the metal/dielectric interface is straightforward \cite{EriksenTserkezisMortensenCox+2024+2741+2751} and is unlikely to qualitatively change the conclusions. 

Below the interband transitions, where the energy of the NPoM gap plasmon mode lies, our results quantitatively reproduce those obtained from the semi-classical specular reflection model (SRM, Fig.~\ref{fig:theory}b), which have been shown 
to match the results of the quantum-mechanical calculations with {\it ad hoc} carefully-constructed atomic layer potentials \cite{RodriguezEcharri2021}. Above the interband transition threshold the calculations diverge since our free-electron model only accounts for the screening from the d-band.
While still remaining a simple one-dimensional model, this calculation represents
the first quantitative self-consistent treatment of optical non-locality at the gold surface in the experimentally-relevant domain.

When the electron temperature increases, the damping modulation and broadening of the interband transitions modifies the contribution of bound electrons to $\varepsilon$ \cite{PhysRevB.61.16956, Karaman2024NCom}. At low frequencies, $\varepsilon$ changes due to the Kramers-Kronig compliance and fractional Umklapp electron-electron scattering \cite{PhysRevLett.81.922, PhysRevB.61.16956, PhysRevB.69.195416}. The increase in the real part introduces corrections in both static and dynamic screening within our model, thus emphasizing the "spill-in" behaviour. The surge of the damping rate (and correspondingly, $\operatorname{Im}~\varepsilon$) modifies the resonance broadening at 2.5~eV. 
Since the shift of the surface plasmon to 2.5~eV from $0.8~\omega_p \approx~7.2$~eV is determined by d-band screening, the modification of $d_\perp$ due to the electron heating is most pronounced in this region. However, a significant tail extends towards lower frequencies, with $\approx 10$\% variations observed around the typical gap plasmon mode energy (1.5~eV). The developed microscopic theory thus predicts that in the "surface plasmon" frequency range ($\sim 1-2$~eV) $\operatorname{Re}~d_{\perp}$ in gold becomes more negative with temperature, indicating that hot electrons exhibit stronger spill-in.

Generally, nonlocal corrections in plasmonic nanoobjects of size $R$ are considered small together with the $d_{\perp}/R$ ratio \cite{Raza2015,Goncalves2020} whereas much stronger ($\sim d_{\perp}/h$) nonlocality can be expected in narrow ($h\sim1$~nm) nanogaps. Numerical calculations indeed demonstrate a strong impact of nonlocality in narrow-gap plasmonic nanoparticles-on-mirror (NPoMs)\cite{Yan2024}. However, low damage threshold of plasmonic nanocavities and strong (up to $10^3$) field localisation and enhancement in the gap\cite{Baumberg2019Review} makes the time-domain experiments with ultrafast lasers extremely challenging. This is further exacerbated by the somewhat malleable nature of gold producing flares \cite{doi:10.1021/acs.nanolett.3c02537} and picocavities \cite{doi:10.1126/science.aah5243} in strong electric fields.

\begin{figure}[ht]
    \centering
    \includegraphics[width=\linewidth]{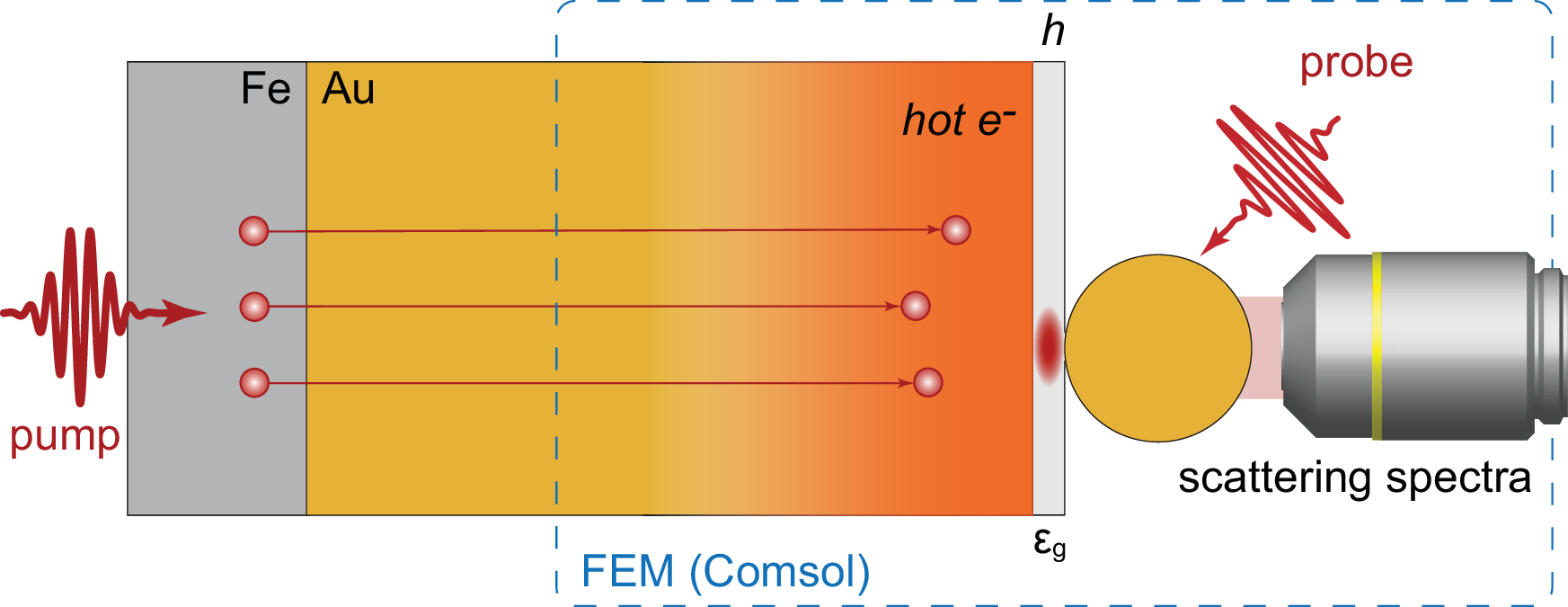}
    \caption{Proposed experimental configuration. An ultrashort pump pulse is absorbed in the Fe layer, inducing hot electron injection from Fe into Au and their subsequent ballistic transport across the Au layer\cite{Melnikov2011}. The delayed probe pulse monitors transient variations of the plasmonic (glowing red hotspot) spectral response of a NPoM nanocavity. The latter is formed by a spherical Au nanoparticle resting on a dielectric layer with a thickness $h$ and dielectric function $\varepsilon_g$. The blue dashed line indicates the region simulated numerically using the Finite Elements Method (FEM) with the help of Comsol Multiphysics.}
    \label{fig:figexp}
\end{figure}

The problem of direct optical heating of the fragile nanoparticle can be avoided in 
our proposed experimental system (Fig.~\ref{fig:figexp}) based on the NPoM geometry with a thick (50-100~nm) gold film epitaxially grown on top of a thin (5-10~nm) layer of iron on a transparent (MgO) substrate \cite{Melnikov2011,AlekhinJPCM2019,KuhnePRR22,Avdizhiyan2024}. A nanocavity is formed by a spherical gold particle placed over the gold surface (mirror). Such nanocavities can be readily produced via chemical means by drop-casting nanoparticle solutions on gold surfaces coated with self-assembled molecular monolayers \cite{doi:10.1126/science.1224823}. Using alkane-thiols \cite{doi:10.1126/science.1224823} or cucurubiturils \cite{doi:10.1021/acs.jpclett.5b02535},
reproducible nanogaps with thicknesses down to 1~nm can be fabricated. Larger gaps can be routinely produced with longer-chain molecules \cite{Li2021SHGNPoM} or atomic-layer-deposited oxide films \cite{Yang2019General}, whereas even thinner spacers ($\sim0.4$~nm) are also feasible with the use of 2D materials \cite{Chen2023}.

In the time-resolved experiments, the Fe emitter is irradiated with an ultrashort laser pulse. 
The electronic transmission of a high-quality Fe/Au interface exhibits a strong dependence on electron energy\cite{Alekhin2017}, which results in the efficient injection of laser-excited hot electrons from Fe into Au. Owing to the rapid thermalization of hot electrons in Fe, such Fe/Au bilayer is capable of generating sub-100 fs bursts of highly energetic ballistic electron propagating though Au at the Fermi velocity $v_F\sim1$~nm/fs. Due to their large ballistic length ($\sim100$~nm)\cite{AlekhinJPCM2019} they traverse the gold layer, modulating its optical properties at the sub-ps timescale. This approach allowed studying ultrafast transport of spin-polarized hot carriers in gold over 50-100 nm\cite{Melnikov2011,Alekhin2017,Razdolski2017,KuhnePRR22,MelnikovPRB22}, significantly exceeding the optical penetration depth ($\approx$ 20 nm), while recent spectroscopic experiments suggested transient variations of optical non-locality in plasmonic gratings \cite{Avdizhiyan2024}.

This experimental approach enables transient non-equilibrium electronic configurations in the proposed NPoM geometry where electronic asymmetry between the mirror and the particle is introduced. The highly energetic electrons reaching the front surface of the gold mirror modulate $\varepsilon_{\rm Au}$ \cite{AlekhinJPCM2019} and, potentially, the spill-out effects at the interface\cite{Avdizhiyan2024} while leaving the nanoparticle unperturbed. The dynamics of electron spill-out and other relevant quantum phenomena can be monitored through the transient dark-field scattering of the NPoM plasmonic nanocavities. As an important practical advantage, our approach enables strong perturbations of the electronic density without an optical damage to NPoM since the probe light
can be made relatively weak.

The proposed hot electron-driven mechanism might not be the only one responsible for the ultrafast modulation of the optical nonlocality in these experimental conditions. Variations in electron density, changes in the dielectric constant as a result of molecular motion in the gap, chemical transformations, and surface charging can all contribute to control both classical and quantum plasmonic phenomena at the ultrafast timescale. Moreover, the case of elevated electron temperature considered here thermodynamically refers to the situation where the excited electronic subsystem has already reached the Fermi-Dirac equilibrium at $T_e$ but thermalization with the cold lattice has not started yet. In typical plasmonic metals this state can be expected at a few hundred fs delay after the exposure to a short ($\lesssim50$~fs) excitation pulse. We thus cannot rule out stronger modulation of the plasmon resonance frequency at shorter time delays when the electronic state is non-thermal.

An inherent connection between the electronic temperature and the non-locality of the optical properties at a metal-dielectric interface enables quantification of the resulting variations of the optical response. Due to the complexity of joining numerical electrodynamics with the quantum-mechanical calculations in the excited state, we consider $d_{\perp}$ as an independent control parameter in numerical simulations. Its dynamic nature is justified with the microscopic model above. 

Using the finite elements method implemented in COMSOL Multiphysics, we calculated dark-field scattering spectra of gold NPoMs with mesoscopic boundary conditions \cite{Yang2019General}.
The NPoM configuration is defined by specifying the nanosphere radius (50~nm), gap thickness (1--5~nm) and dielectric function ($\varepsilon_g=2$).
With this setup, COMSOL routines enable the extraction of key optical observables, including scattering cross-sections, absorption spectra, and near-field electric field enhancements within the simulated region.

Incorporating the Feibelman approach, the mesoscopic boundary conditions were implemented in the COMSOL model through weak-form integrals within the scattering formulation of Maxwell’s equations\cite{Yang2019General}. In this work we expand the reported two-dimensional model\cite{Yang2019General} into the three-dimensional domain. Numerical tests demonstrated excellent continuity of our 3D simulation model in the $d\rightarrow0$ limit, yielding a perfect agreement with the local results 
(Fig.~S1).

\begin{figure}[ht]
    \centering
    \includegraphics[width=\linewidth]{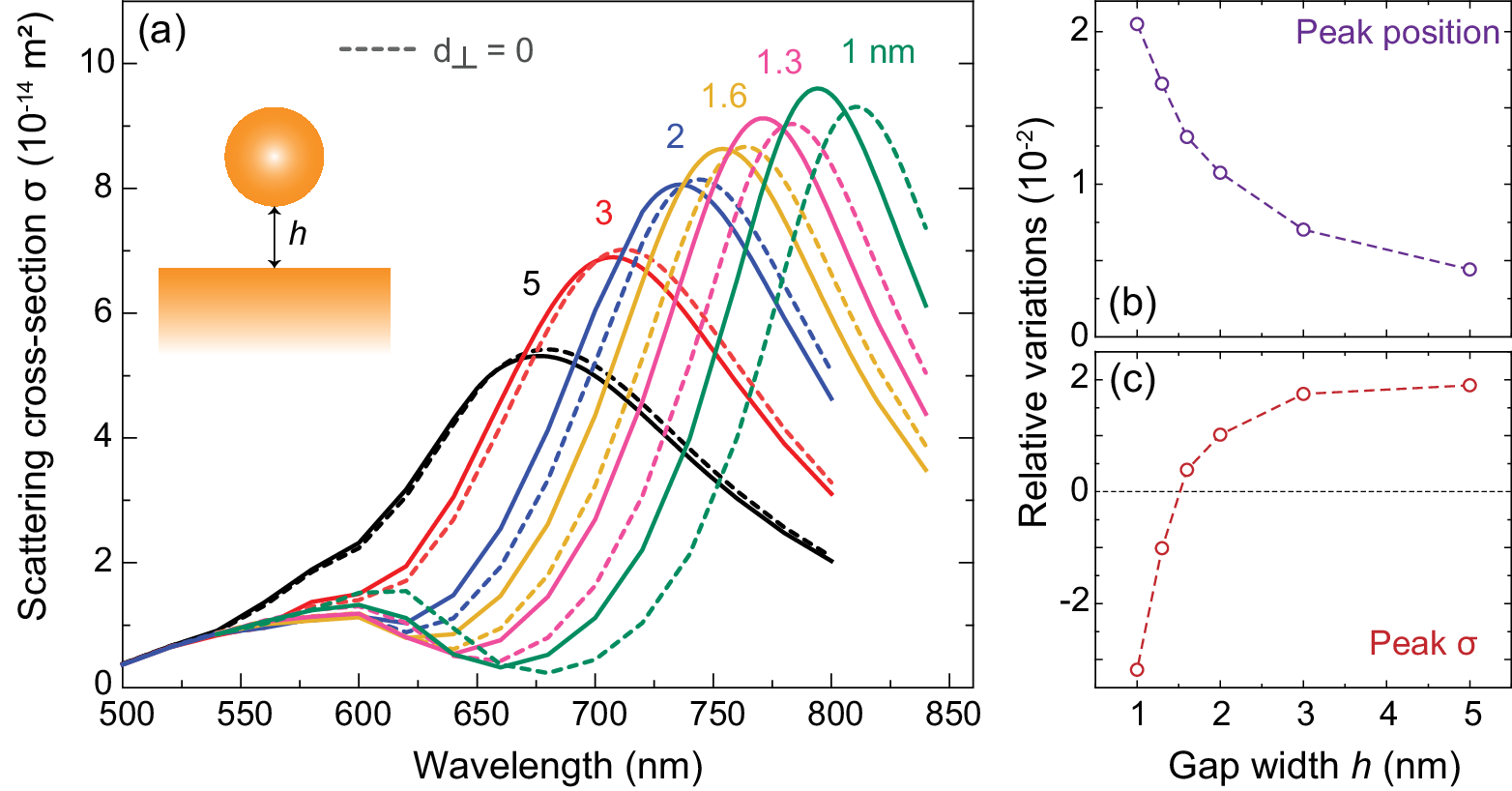}
    \caption{Role of nonlocality and gap size in gold NPoM: (a) Scattering cross-sections of NPoMs with varied gap width in local theory (dashed lines) and nonlocal theory with d$_{\perp}$ taken from Ref.~\cite{RodriguezEcharri2021} (solid lines); (b-c) Relative variations in peak position $\Delta\lambda/\lambda$ (b) and scattering amplitude at the maximum $\Delta\sigma/\sigma$ (c) between the local and nonlocal theories.}
    \label{fig:figgap}
\end{figure}

Figure~\ref{fig:figgap} shows scattering spectra of NPoMs where, upon reducing the gap width $h$ from 5 to 1~nm, the resonance redshifts\cite{doi:10.1126/science.1224823}.
This trend is reproduced in both non-local ($d_{\perp}\neq0$, solid lines)  and local ($d_{\perp}=0$, dashed lines) calculations. The impact of non-locality increases toward smaller gaps (Fig.~\ref{fig:figgap}b): NPoMs with $h=1$~nm demonstrate the strongest blueshift when $d_{\perp}\neq0$ is introduced and thus are the most promising for active modulation. Although NPoMs with even smaller gaps might exhibit larger effects, in practice the gap thickness uniformity becomes a challenge. Furthermore, since at $h\lesssim0.5-1$~nm electron tunnelling becomes important\cite{Savage2012,Zhu2016}, we focus on the 1~nm gap geometry.

\begin{figure}[ht]
    \centering
    \includegraphics[width=\columnwidth]{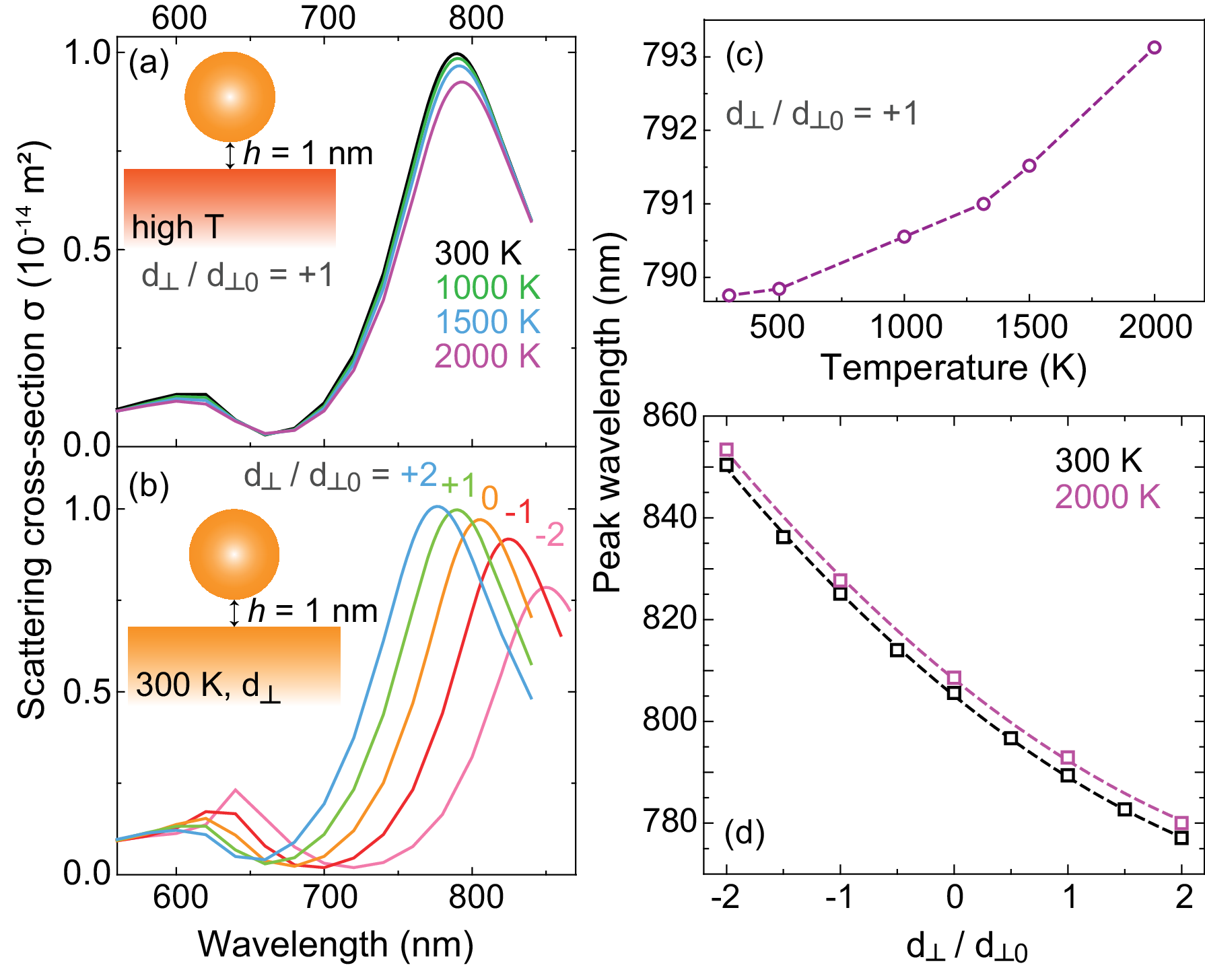}
    \caption{Interplay of nonlocality and substrate heating in Au NPoMs with the gap width $h=1$~nm. 
    (a-b) Scattering cross-section spectra of NPoMs with: (a) varied mirror temperatures and $d_{\perp}=d_{\perp0}$, and (b) cold mirror at 300 K and varied $d_{\perp}$. 
    (c-d) Extracted scattering peak positions at various mirror temperatures and $d_{\perp}/d_{\perp 0}=1$ (c), and non-locality $d_{\perp}$ in the mirror at $T_e=300$~K and $2000$~K (d). 
    Because $d_{\perp 0}<0$, the horizontal axis in (d) points in the direction of further decrease of $d_{\perp}$, i.e. stronger spill-in.
    }
    \label{fig:figtemp}
\end{figure}

To set up a reference point for the non-locality, we employ $d_{\perp 0}\approx-0.48$~{\AA} obtained\cite{RodriguezEcharri2021} for gold at $\lambda=790$~nm, i.e. in the vicinity of the gap plasmon mode at $h=1$~nm. 
The impact of the electronic temperature in the mirror $T_e$ and non-locality $d_{\perp}$ is seen in Fig.~\ref{fig:figtemp}a-b. The resonant wavelength exhibits much stronger sensitivity to the Feibelman non-locality than the linewidth (cf.~Fig.~S6). Naturally, this is the result of our assumption $\operatorname{Im}d_{\perp}=0$ justified by the microscopic description below the onset of interband transitions. Both temperature rise and an increase of the electron spill-out 
redshift the gap mode. 
The 
extracted peak positions (Fig.~\ref{fig:figtemp}d, dashed lines) exhibit a temperature-independent slope of $\partial\lambda/\partial d_{\perp}\approx290$. 
Interestingly, a second order non-local correction $\partial^2\lambda/\partial d_{\perp}^2\approx1800$~nm$^{-1}$ is also discernible from our results (Fig.~S4). To our knowledge, this is the first identification of the second-order Feibelman-like contribution
, indicating exceptional susceptibility of the gap plasmons in non-equilibrium quantum plasmonics.

It is instructive to incorporate NPoM nonlocal corrections into a simple analytic theory capable of replicating the numerical results qualitatively. Recall that the gap mode is the ultimate case for the hybridized surface plasmons at the two adjacent interfaces \cite{Baumberg2019Review}. 
Resonant plasmonic nanocavities can be modelled as effective electric circuits at optical frequencies \cite{Benz:15}, where NPoM is replaced with a nanoparticle dimer for convenience.
This coupled oscillators model exhibits two resonances corresponding to the LSPR of the individual nanoparticles and the gap mode, respectively. The latter is enabled by the capacitive coupling $C_g$ between the $LC$-circuits on the particles.

\begin{figure}[ht!]
    \centering
    \includegraphics[width=\linewidth]{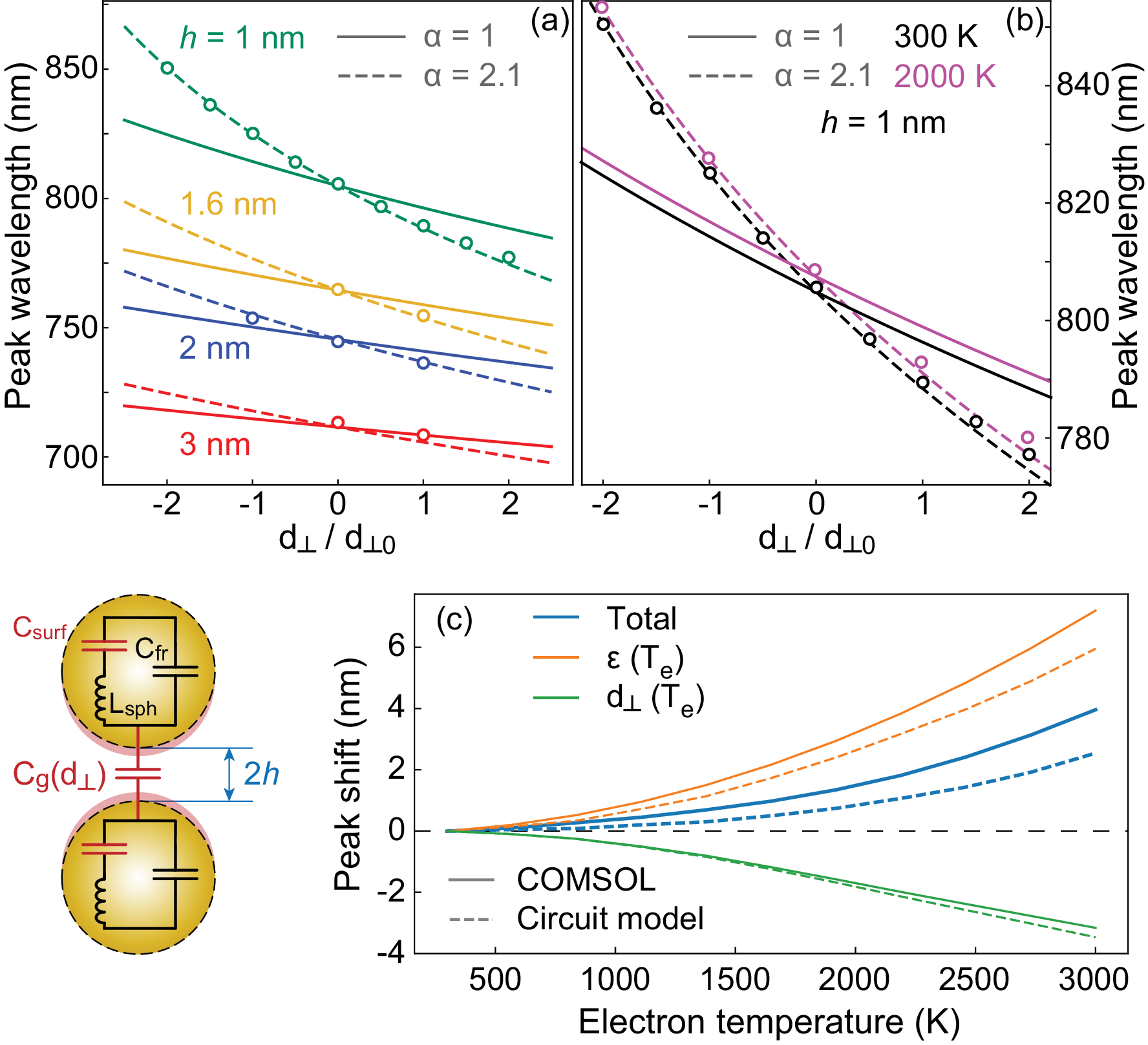}
    \caption{
    Variations of the calculated resonant wavelengths with the Feibelman parameter $d_{\perp}$ for different gap widths (a) and temperatures of the gold mirror (b) with the analytic circuit model. The dashed and solid lines represent the expected scaling of the gap capacitance with $d_{\perp}$ with and without the correction factor $\alpha = 2.1$, respectively. The inset in the bottom-left corner shows a schematic circuit of a nanoparticle dimer with non-locality. The red-colored $C_{\rm g}$ is modified through the Feibelman non-locality $d_{\perp}$ which can be described by the effective gap width renormalization (Eq.~\ref{eq:circuit:capac}, red-shaded spheres).
    (c) Calculated resonance shifts with temperature: the impact of bulk $\varepsilon$ (orange) and Feibelman non-locality $d_{\perp}$ (green) with temperature, and the total shift (blue). The solid and dashed lines indicate the results of numerical simulations and the circuit model, respectively.
    }
    \label{fig:circuit}
\end{figure}

Adapting the model to capture the nonlocal response, we need to account for both $d_{\perp}/R$ and $d_{\perp}/h$ corrections.
The former is realized with a surface capacitance $C_{\rm surf}$ (Supporting Information S1.1, Fig.~\ref{fig:circuit}a). Employing the Drude model with $\varepsilon_{\infty}$ and $\omega_p$, the resonance wavelength of the gap plasmon is:

\begin{equation}\label{eq:circuit:res}
\lambda_{0} = \lambda_p\sqrt{\varepsilon_{\infty} + 2\varepsilon_h(1 + 2\eta)\frac{1 + d_{\perp}/R}{1 - 2d_{\perp}/R}},
\end{equation}
where $\eta = 2C_g/(\varepsilon_h R)$ is the ratio of the gap and sphere capacitances, $\varepsilon_h=1$ is the host medium dielectric function and $\lambda_p=2\pi c/\omega_p$.
%

In conventional NPoM geometries, $R\gg h$, thus emphasizing the $\sim d_{\perp}/h$ corrections. In Eq.~\ref{eq:circuit:res}, they are responsible for the modification of the effective gap width $h$ thereby adjusting $\eta$ (Supporting information S1.1.3):

\begin{equation}\label{eq:circuit:capac}
\eta = \frac{n_g^{\chi}}{\varepsilon_h}\ln\left(1 + \frac{R}{2(h - 2\alpha d_{\perp})}\theta_{max}^2\right).
\end{equation}
A phenomenological parameter $\alpha$ serves to compensate the model's underestimation of the role of $d_{\perp}$ due to local assumptions underlying the derivation of Eq.~\ref{eq:circuit:capac}. Because the physical meaning of $d_{\perp}$ shows the displacement of the induced electronic density from the (sharp) interfaces, this signifies a tempting approach to introduce an effective renormalization of the gap width due to the spill-in (or -out) of the electron density.
Within the model, $\varepsilon_{\infty}$ and $\chi$ are used to match the resonant frequencies of nanocavities for $d_{\perp}=0$. We estimated $\varepsilon_{\infty}$ from the LSPR scaling in water (Fig.~S1), incorporating subtle retardation effects\cite{Benz:15}. The role of $T_e$ can be accounted for by calculating the $\operatorname{Re}\varepsilon_{Au}$ variations using the microscopic theory (Eqs.~\ref{eq:rosei:eps}--\ref{eq:rosei:gamma}) whereas we corrected $\varepsilon_{\infty}$ by $\Delta\varepsilon = \operatorname{Re}~\varepsilon(2000~K) - \operatorname{Re}~\varepsilon(300~K) \approx 0.22$.
The base theory (Fig.~\ref{fig:circuit}a-b, $\alpha = 1$) severely underestimates the role of nonlocality. Yet, an excellent agreement of the analytic theory with the numerically computed resonance wavelengths at both $T_e$ and various gap widths is obtained by setting $\alpha=2.1$,
even though the dimer-based circuit model does not directly incorporate the heated mirror-induced asymmetry. Moreover, the model is capable of replicating the nonlocal behaviour in the wide $h-d_{\perp}$ parameter space (Fig.~\ref{fig:circuit},a). Supporting our numerical treatment, this result introduces an intuitive analytic way to model nonlocal plasmonic nanocavities out of thermal equilibrium. 

The need for a correcting coefficient $\alpha$ can also be inferred from comparing the sensitivity of the gap plasmon wavelength $\lambda_0$ to the variations of $h$ and $d_{\perp}$ in numerical simulations. In the vicinity of $h=1$~nm (Fig.~\ref{fig:figgap}), we find $\partial\lambda_0/\partial h\approx-118$. Recalling the variations of the resonance position with $d_{\perp}$ from Fig.~\ref{fig:figtemp}d, for the relative sensitivity we obtain $\xi=(\partial\lambda_0/\partial h)/(\partial\lambda_0/\partial d_{\perp})\approx -0.41$. Although at smaller gaps $|\partial\lambda_0/\partial h|$ surges, $\xi>-1$ still holds in the practical range of $h$, indicating a more complicated gap renormalization in the presence of non-local plasmonic response.

We can now illustrate the heating-induced shifts of the gap plasmon resonances as predicted by both analytic model and numerical simulations. In both cases, we calculate contributions from the temperature-induced variations of the bulk $\varepsilon$ and non-locality $d_{\perp}$ as well as the total shifts (Fig.~\ref{fig:circuit}c). Although the analytic circuit model suffers from multiple assumptions and simplistic parametrization, it yields a good qualitative agreement with the numerical results.

Both approaches show that the $d_{\perp}(T_e)$ dependence significantly reduces the resonance shift at elevated temperatures ($\sim50\%$), enabled by the opposite signs of the $\varepsilon(T_e)$ and $d_{\perp}(T_e)$ contributions.
Whereas the former dependence is well established, the latter provides important insights into the non-equilibrium non-locality of the optical response of nanoplasmonic systems with noble metals. 
The calculated reduction of the heating-induced resonance shifts is directly observable in the experiment, {\it e.g.} with ultrafast laser-induced injection of hot electrons (Fig.~\ref{fig:figexp}). The $\varepsilon$-driven contribution can be estimated in other plasmonic systems where the role of non-locality is suppressed. This work thus provides a comprehensive framework for the active, ultrafast non-local plasmonics.

In conclusion, we calculated thermal variations of the Feibelman nonlocality in gold and proposed an experimental approach to actively control non-classical plasmonic phenomena in nanogaps with ultrafast laser-induced hot electron bursts. 
Our analytical and numerical results indicate sizeable shifts of a fundamental gap plasmon mode in NPoM nanocavities, that could be monitored with state-of-the-art ultrafast spectroscopy.
Minimizing the optical damage to the nanocavities, this approach facilitates the development of active quantum plasmonics and other fundamental studies of high-energy and non-equilibrium phenomena. It might prove vital in studying coherent molecular motion in nanogaps \cite{txdw-nqvn}, picocavities \cite{doi:10.1126/science.aah5243} and dynamic strong coupling \cite{Hu2024} as well as probing hot carrier dynamics with atomic resolution \cite{Luo2025}. 
The asymmetric nature of the predicted modification is interesting for the fundamental studies of nanocavities with broken symmetry, {\it i.e.} made of a combination of "spill-in" and "spill-out" metals \cite{Yang2019General} such as recently developed robust magnesium nanoparticles\cite{BigginsNanoLett2018} on gold mirrors. Furthermore, a feasible integration of the proposed geometry with 2D materials opens the door to the fascinating perspectives of dynamic control of electron tunneling \cite{Savage2012}. Taking advantage of the single-molecule sensitivity of NPoM plasmonic cavities, our proposal holds promise for advance studies of hot carrier-driven chemistry and ultrafast catalysis in nanogaps.




\subsection*{Funding}
This research is partially supported by Universiti Malaya Research Excellence Grant 2025 (Project No.~UMReG034-2025).


\section*{Acknowledgements}
The authors thank Prof. Anatoly~V.~Zayats and Prof. Jeremy~J.~Baumberg for reading the initial version of the manuscript and providing helpful suggestions. 


\newpage

\section {Supporting Information}

\section{Theory}

\subsection{Nonlocal circuit model for NPoM}

To implement a simple analytic model that would qualitatively predict the nonlocal response of a NPoM plasmonic cavity, we adapt the technique pioneered in \cite{PhysRevLett.95.095504} treating nanoparticles as nanoscale curcuit elements at optical frequencies.

\subsubsection{Isolated plasmonic nanoparticle}

The starting point is the derivation of the electric field of an individual metallic sphere described by the dielectric function $\varepsilon_m (\omega)$, in the dielectric environment with $\varepsilon_d$ in a uniform external electric field. For this we recall the general solution of a Laplace equation in spherical coordinate system:

\begin{equation}
\phi(r, \theta) = \sum_{l = 0}^{\infty}\left(A_lr^l + \frac{B_l}{r^{l+1}}\right)P_l(\cos \theta),
\end{equation}
where $P_l$ are Legendre polynomials. To retain the simplicity and since we are only interested in the main dipole feature of the scattering spectrum, we restrict ourselves to the simplest quasistatic approximation and, assuming the external field $\textbf{E}_0 || \textbf{z}$, seek the electrostatic potential in the form:
\begin{align}
    \phi_{in}(r, \theta) = Ar\cos\theta, \nonumber \\
     \phi_{out}(r, \theta) = E_0 r\cos\theta + \frac{B}{r^2}\cos\theta,
\end{align}
Applying the standard boundary conditions -- continuity of the potential and radial displacement $D_r$ at $r = R$ (radius of the nanoparticle) -- yields the textbook solution for this problem. Extension of this approach to include nonlocal corrections within the Feibelman framework is straightforward, and involves replacing continuity of the potential with the new boundary condition:

\begin{equation}
  E^m_{\theta} - E^d_{\theta} = -d_{\perp}\nabla_{\|}\left(E^m_{r} - E^d_{r}\right)
\end{equation}

This condition, after straightforward algebra, yields the following result:

\begin{align}
    \textbf{E}_{in} = \frac{3\varepsilon_h}{\varepsilon_m + 2\varepsilon_h - 2 d_{\perp}\left(\varepsilon_m - \varepsilon_h\right)/R}\cdot \textbf{z}, \nonumber \\
     \textbf{E}_{out} = E_0\cdot \textbf{z} + \frac{3\textbf{r}(\textbf{p}\cdot\textbf{r}) - \textbf{p}}{\varepsilon_h R^3},
\end{align}

where $\varepsilon_m$ and $\varepsilon_h$ are the metal and host medium dielectric functions, and the nonlocal dipole moment of the nanoparticle is given by:
\begin{equation}
    (\textbf{p}\cdot \textbf{z}) = \varepsilon_h R^3\frac{\varepsilon_m - \varepsilon_h + d_{\perp}(\varepsilon_m - \varepsilon_h)/R}{\varepsilon_m + 2\varepsilon_h - 2 d_{\perp}\left(\varepsilon_m - \varepsilon_h\right)/R}
\end{equation}

It is clear that these expressions retain the overall structure of the local solution, including a representation of the field outside the particle as a sum of the external and dipole fields, and converge to the known answer in the absence of nonlocality. We would also like to point out that nonlocal polarisabilities could be similarly derived for an arbitrary order in the multipolar expansion \cite{Gonçalves2020}. 

With the expression of the fields known, we can adapt the circuit model from Ref.~\cite{PhysRevLett.95.095504} to our problem. Identically, the continuity of $D_r$ allows to formulate the "effective" Kirchhoff current law for the system:

\begin{equation}
    -i\pi\omega(\varepsilon_h - \varepsilon_m) R^2 E_0 = -2i\pi\omega\varepsilon_h R^2 E_0 \left(1 + \frac{d_{\perp}}{R}\right)\Gamma - i\pi\omega\varepsilon_m R^2 E_0 \left(1 - \frac{2d_{\perp}}{R}\right)\Gamma,
\end{equation}
with $\Gamma = (\varepsilon_m - \varepsilon_d)/(\varepsilon_m + 2\varepsilon_h - 2d_{\perp}(\varepsilon_m - \varepsilon_h)/R)$. As in Ref.~\cite{PhysRevLett.95.095504}, we interpret the first and second terms on the right as fringe and sphere current respectively, however due to boundary condition (S3) introducing a discontinuity of the potential we can no longer satisfy the Kirchhoff voltage law with just two circuit elements. Instead, we introduce a new element with the impedance Z$_{surf}$, which is connected in series with the Z$_{sphere}$, and serves to incorporate a potential jump due to the additional dipole layer at the interface. After computing the average potential difference between the upper and lower halves of the sphere we can express the impedances of all three circuit elements as follows:

\begin{align}
Z_{\rm sphere} = (-i\pi\omega\varepsilon_m R)^{-1}, && Z_{\rm fringe} = (-2i\pi\omega\varepsilon_h R)^{-1}, && Z_{\rm surf} = \frac{3d_{\perp}}{-i\pi\omega\varepsilon_h R^2}\left(1 - \frac{2d_{\perp}}{R}\right)^{-1}
\end{align}

This result allows a clear interpretation of the role of nonlocality. For the noble metals at low frequencies, where $d_{\perp}$ is real and negative, this circuit element represents an additional capacitance (or negative inductance) connected in series to the classical inductance of the nanoparticle. It therefore serves to increase the resonance frequency of the corresponding LC circuit, as reported elsewhere\cite{ChristensenPRL2017}.

\subsubsection{Coupled circuit model for the NPoM}

The developed non-local circuit model allows us to explore the coupled structures with ease. As in the previous works \cite{Benz:15} we treat the nanoparticle-over-mirror as a dimer produced by coupling of the nanoparticle plasmon
with its image charges. Following previous works we model this coupling by introducing an LRC circuit where $L_g$, $C_g$ and $R_g$ represent the effective inductance, capacitance and resistance of the gap, respectively. The total impedance of the system $Z_{\rm total}$ is then given by:

\begin{equation}
    Z_{\rm total} = \frac{2\left(1 + d_{\perp}/R\right)}{-i\pi\omega R\left(\varepsilon_m + 2\varepsilon_h - 2d_{\perp}(\varepsilon_m - \varepsilon_h)/R\right)} + \frac{1}{-i\omega C_g + [R_g - i\omega L_g]^{-1}}
\end{equation}
Assuming for the sake of simplicity that the dielectric function of metal is represented by the lossless Drude model and parametrised by the plasma frequency $\omega_p$ and $\varepsilon_{\infty}$ we can obtain the resonant frequency from the poles of $\operatorname{Im}\left(Z_{\rm total}\right)$. The solutions could proceed using the same algorithm as outlined in Ref.~\cite{Benz:15}. However here we note that a dipolar plasmon frequency for our system, that assumes a dielectric spacer, can be obtained by scaling $\varepsilon_{\infty}$ and $\omega_p^2$ by $\frac{1 - 2d_{\perp}/R}{1 + d_{\perp}/R}$. For the resonant plasmon wavelength $\lambda_0$ we obtain:

\begin{equation}
\lambda_{0} = \lambda_p\sqrt{\varepsilon_{\infty} + 2\varepsilon_h(1 + 2\eta)\frac{1 + d_{\perp}/R}{1 - 2d_{\perp}/R}},
\end{equation}
where $\lambda_p$ is the plasma wavelength, $\lambda_p=2\pi c/\omega_p$, and $\eta = C_g/C_{\rm fringe}$ is the normalised gap capacitance.

\subsubsection{Gap capacitance}

While Eq.~S7 allows to estimate the role of nonlocality of the individual nanoparticles they clearly introduce corrections of the order of $d_{\perp}/R$ and therefore fail to incorporate the leading role of a small gap in the observable effect. To incorporate the latter we need to consider the change in gap capacitance induced by nonlocality. The simplest model of such kind would involve
an adjustment of the gap thickness $h$ by a nonlocal correction:


\begin{equation}
\eta = \frac{n_g^{\chi}}{\varepsilon_h}\ln\left(1 + \frac{R}{2(h - 2\alpha d_{\perp})}\theta_{max}^2\right)
\end{equation}

Physically this represents an additional contribution of $\alpha d_{\perp}E$ to the potential difference across the capacitor, that arises from potential discontinuities at the electrodes (for $\alpha = 1$). It is important to note that the in-plane gradient in Eq.~S3 effectively violates the base assumption of the model - representing the field of each subdivision of curved electrode as equivalent to field of the dihedral capacitor of the same size and orientation \cite{Hudlet1998}. The unaccounted for contribution of the nearby points is what potentially makes the model to underestimate the nonlocality in a given setting. To compensate for this inconsistency we introduce the phenomenological parameter $\alpha$ to scale the contribution up to the observable values. 

\begin{figure}[h]
    \centering
    \includegraphics[width=0.6\linewidth]{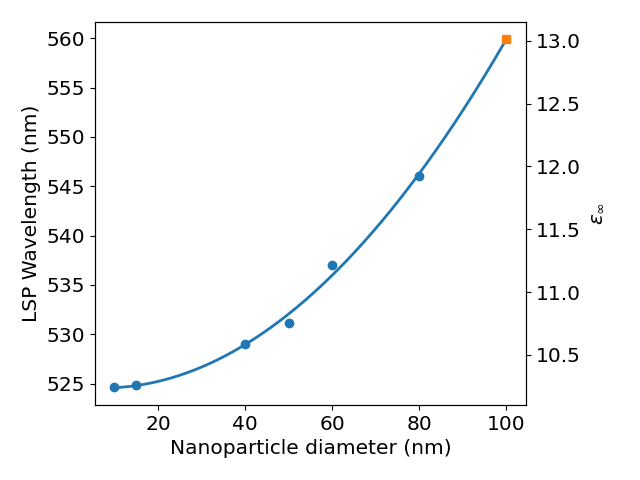}
    \caption{LSP resonances of spherical gold nanoparticles in water and recovered $\varepsilon_{\infty}$. Data from Ref.~\cite{Benz:15}. The solid line is a phenomenological polynomial extrapolation of the data. The orange square represents the nanoparticle diameter used in the present work (100~nm).}
    \label{sfig:lsp}
\end{figure}

The value $\chi$, theoretically expected to be close to 2, was used in previous works \cite{Benz:15} as a fine-tuning parameter to adjust the position of the NPoM resonance (in the local case, at $d_{\perp} = 0$), acting together with the size-dependent $\varepsilon_{\infty}$. Here we follow the same recipe and obtain the unknown high-frequency dielectric constant $\varepsilon_{\infty}\approx~13.0$ by extrapolating the dataset \cite{Benz:15} of a single-particle localised surface plasmon resonances in water (Figure.~\ref{sfig:lsp}). Subsequently, using the value $\chi = 2.02$ we obtain an excellent agreement of the wavelength of the fundamental NPoM resonance (805.6~nm) with the numerical simulations (804.9~nm). Moreover, estimating the thermal change of $\varepsilon_{\infty}$ from our microscopic model (Section~S1.2) to be $\approx~0.22$ upon heating to 2000~K, we are able to match the resonance positions at elevated electron temperatures without any extra free parameters.

\begin{figure}[h!]
    \centering
    \includegraphics[width=0.6\linewidth]{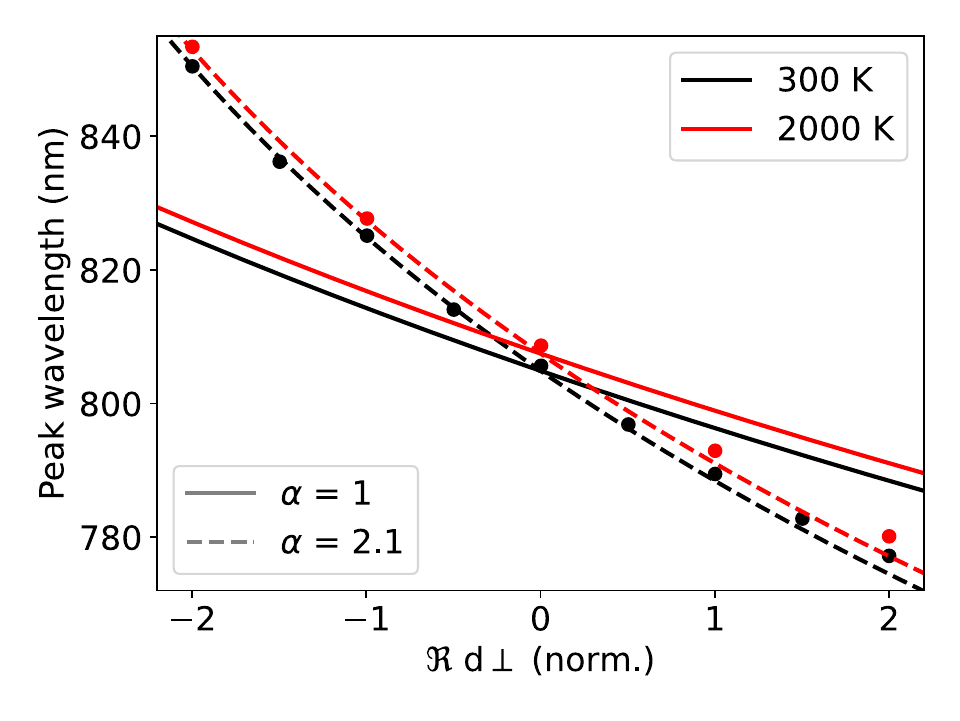}
    \caption{Positions of the NPoM resonance obtained from the nonlocal circuit model with $\alpha$ = 1 (solid) and 2.1 (dashed), at the temperatures 300 K (black) and 2000 K (red). The points are the same as Fig.~4d in the main text.}
    \label{sfig:circuit}
\end{figure}

\subsection{Time-depended DFT}

TD-DFT calculations used in this study follow the route that is well established in the literature \cite{Liebsch1997} and incorporate three main approximations - "jellium" treatment of positive ionic background, adiabatic local density  approximation for the exchange-correlation potential and semi-classical approximation of the screening imposed by the d-electrons, known as the s-d model \cite{PhysRevLett.71.145, PhysRevB.52.14219}. For consistency with the established body of literature and for simplicity we also used Wigner interpolation for the exchange correlation energy. Although more accurate, newer expressions are available and frequently used in the literature \cite{RodriguezEcharri2021} we did not notice a significant deviation of the results based on the exchange correlation approximation used. 

The main deviation of our approach from the previous works is the use of microscopic temperature-dependent dielectric function of gold to devise the contribution of the d-band screening. The imaginary part of the interband dielectric function of gold, as stated in the main text is given by:

\begin{equation} \label{seq:rosei:eps}
    \operatorname{Im}[\varepsilon_{d}(\omega, T_e)] = (\hbar\omega)^{-2}[A_X J_{X}(\omega, T_e)+ A_{L_4^+} J_{L_4^+}(\omega, T_e)+A_{L_{5+6}^+} J_{L^+_{5+6}}(\omega, T_e)],
\end{equation}

where the joint densities of states used are calculated from the known band parameters through the energy distribution of the joint density of states:

\begin{equation} \label{seq:integral}
    J_{i}(\omega,t)=\int_{E_{min}}^{E_{max}}D_i(E,\omega) f(E, T_e, t)dE,
\end{equation}

with $f$ representing the Fermi-Dirac occupation factors, and the band parameters used to calculate $D$ are given in the table \ref{table1}. The weights $A_i$ were chosen to match the room temperature dielectric function of gold (see Figure \ref{sfig:epsilon}). The real part of $\varepsilon$ was obtained by Kramers-Kronig transformation, and summed up with the Drude term, featuring additionally an $\varepsilon_{\infty}$ which is also interpreted as a bound electron contribution and included in the calculation of the screening dielectric function, as discussed in the main text. 

\begin{table}[ht]
\caption{Band-structure parameters near the X- and L-symmetry points used in the model for gold}
\centering
\resizebox{\textwidth}{!}{%
\begin{tabular}{c | c | c | c | c | c | c | c | c | c | c| c | c | c | c | c}
\hline\hline
 &\multicolumn{10}{|c|}{Mass ($m_0$)}&\multicolumn{5}{|c|}{Energy (eV)} \\
\hline \hline
 & $m^X_{p{\parallel}}$ &$m^X_{p{\perp}}$ & $m^X_{d{\parallel}}$ &$m^X_{d{\perp}}$ & $m^L_{p{\parallel}}$ &$m^L_{p{\perp}}$ & $m^{L^+_4}_{d{\parallel}}$ &$m^{L^+_4}_{d{\perp}}$ & $m^{L^+_{5+6}}_{d{\parallel}}$ &$m^{L^+_{5+6}}_{d{\perp}}$ & $\hbar\omega_7^X$ & $\hbar\omega_6^X$ & $\hbar\omega_{4^+}^L$ & $\hbar\omega_{4^-}^L$  & $\hbar\omega_{5+6}^L$\\
\hline 
 & 0.12 & 0.22 & 0.91  & 0.75 & 0.25 & 0.23 & 0.57 &  0.49 & 0.7 & 0.63 & 1.72 & 1.46 & 2.6 & 0.72 & 1.45\\
\hline \hline
\end{tabular}%
}
\label{table1}
\end{table}

The groundstate and induced density calculations proceed as described elsewhere\cite{PhysRevB.52.14219, PhysRevB.36.7378, Liebsch1997, Yang2019General}, using $\varepsilon(0, Te)$ and $\varepsilon(\hbar\omega, T_e)$ in the screened Poisson equation and the long-wavelength approximation of external potential $\psi_{ext} = -2\pi z$. The first order induced density is  then given by:
\begin{equation}
\delta n = \int dz' \chi(z, z', \omega)[\psi_{ext}(z', \omega) \\+ \psi_{ind}(z', \omega) + \delta V_{xc}(z', \omega) ],
\end{equation}

where in adiabatic LDA $\delta V_{xc} = \frac{\delta V_{xc}\left[n(z, \omega)\right]}{\delta n(z, \omega)}$, and the linear response function $\chi$ is constructed from the ground state wave functions for the s-electrons \cite{PhysRevB.36.7378}. 

The dynamic force sum rule used to aid in the evaluation of d$_{\perp}$ in the presence of a polarisable medium has been extensively reported in the literature \cite{PhysRevB.52.14219, Liebsch1997} and is given by:

\begin{align}
d_{\perp} = \frac{\varepsilon - \varepsilon_d}{\varepsilon + 1} d_s 
            &+ \frac{\varepsilon_d - 1}{\varepsilon + 1} d_d, \nonumber \\
d_d = \frac{\varepsilon_d - \varepsilon}{\varepsilon_d} d_s 
       &+ \frac{\varepsilon + 1}{\varepsilon_d} \int_0^{\infty} z \, \delta n(z) \, dz, \nonumber \\
d_s = \frac{\varepsilon - \varepsilon_d}{\varepsilon} \Bigg[
         \varepsilon_d \int_0^{\infty} \delta n(z) \, dz
         &- (\varepsilon_d - 1) \Bigg(
             \frac{\varepsilon_d}{\varepsilon + 1}\int_{-\infty}^0 f(z) \, dz + \int_0^{\infty} z \, \delta n(z) \, dz
             \nonumber \\ &+ \int_{-\infty}^0 f(z) \,dz \int_{-\infty}^z \delta n(z') \, dz'
           \Bigg)
       \Bigg].
\end{align}

with $f = (n(z) - n_{+}(z))/n_0$, $n(z)$ is the groundstate density profile, and $n_{+}$ is the step-like density of ionic background. Alternatively, as reported in previous works, this step can be bypassed by invoking a concept of nonlocal reflection coefficient \cite{RodriguezEcharri2021, Yang2019General}.

In addition, to demonstrate quantitative agreement of our screened calculations within their range of applicability, with the state of the art, we adapted the so called infinite barrier or specular reflection model (SRM) following \cite{RodriguezEcharri2021}. Within this formalism, the expression for the Feibelman parameter is given by:
\begin{equation}
d_{SRM} = -\frac{2}{\pi}\frac{\varepsilon(\omega)}{\varepsilon(\omega) - 1}\int^{\infty}_0\frac{1}{k^2}\left(\frac{1}{\varepsilon_L(\omega, k)} - \frac{1}{\varepsilon(\omega)}\right)dk,
\end{equation}
where $\epsilon_L(\omega, k)$ is an appropriate longitudinal dielectric function of metal. Choosing the hydrodynamics dielectric function with the parameter $\beta = \sqrt{\frac{3}{5}}v_F$ we can perform the integration analytically and obtain: 

\begin{equation}
d_{SRM} = i\frac{\varepsilon\beta}{\varepsilon - 1}\frac{\left(\varepsilon_d/\varepsilon - 1\right)^{\frac{3}{2}}}{\omega_p\sqrt{\varepsilon_d}}.
\end{equation}

This expression was used to produce the dashed curves in Figure~1b in the main text.

\pagebreak

\section{Electromagnetic simulations}

Numerical simulations were performed in COMSOL Multiphysics using the Electromagnetic Waves, Frequency Domain interface in the Wave Optics Module. We implement the mesoscopic boundary conditions in the COMSOL model through weak-form integrals within the scattering formulation of Maxwell’s equations\cite{Yang2019General}.

\begin{gather}
\begin{aligned}
    \nabla\times\mu^{-1}\nabla\times E_{\rm sc} - \varepsilon(r,\omega)k^2 E_{\rm sc} =\Delta\varepsilon(r,\omega)k^2 E_{\rm inc}
    \label{mesobc}
\end{aligned}
\end{gather}
Here $E_{\rm inc}$ and $E_{\rm sc}$ are the incident and scattered fields while $\Delta\varepsilon$ is the permittivity contrast between the scattering object and its background. To improve numerical efficiency and stability, $d_{||}$ was incorporated into the non-classical boundary conditions via a surface current term, while $d_{\perp}$ was treated using an auxiliary potential method ($K$ is an induced in-plane surface current while $E_{sc}^c$ is a field with continuous parallel components\cite{Yang2019General}):
\begin{gather}
\begin{aligned}
K(r)=id_{||}\omega_{||}\lbrack D_{||}\rbrack\\
E_{\rm sc}=E_{\rm sc}^c+\nabla\psi
\end{aligned}
\end{gather}
Further details can be found in Ref.~\cite{Yang2019General}. The two-dimensional model was expanded into the three-dimensional domain. 
The model geometry comprised a rectangular computational domain of a cubic shape with lateral dimensions of 500~nm, enclosed by a 150-nm-thick perfectly matched layer (PML) to ensure ideal absorption outside the active region of the system. The gold mirror layer with a thickness of 50~nm rests on an insulating ($\varepsilon=2.25$) 200~nm-thick layer. The finite-element mesh contained approximately $8\times10^5$ elements, with a minimum element size of 0.6~nm.
In the main text calculations, we set $d_{||}=0$ due to the charge neutrality and only consider variations of $d_{\perp}$. 

\pagebreak
\subsection{Consistency of the non-local computational model}

\begin{figure}[h]
    \centering
    \includegraphics[width=0.48\linewidth]{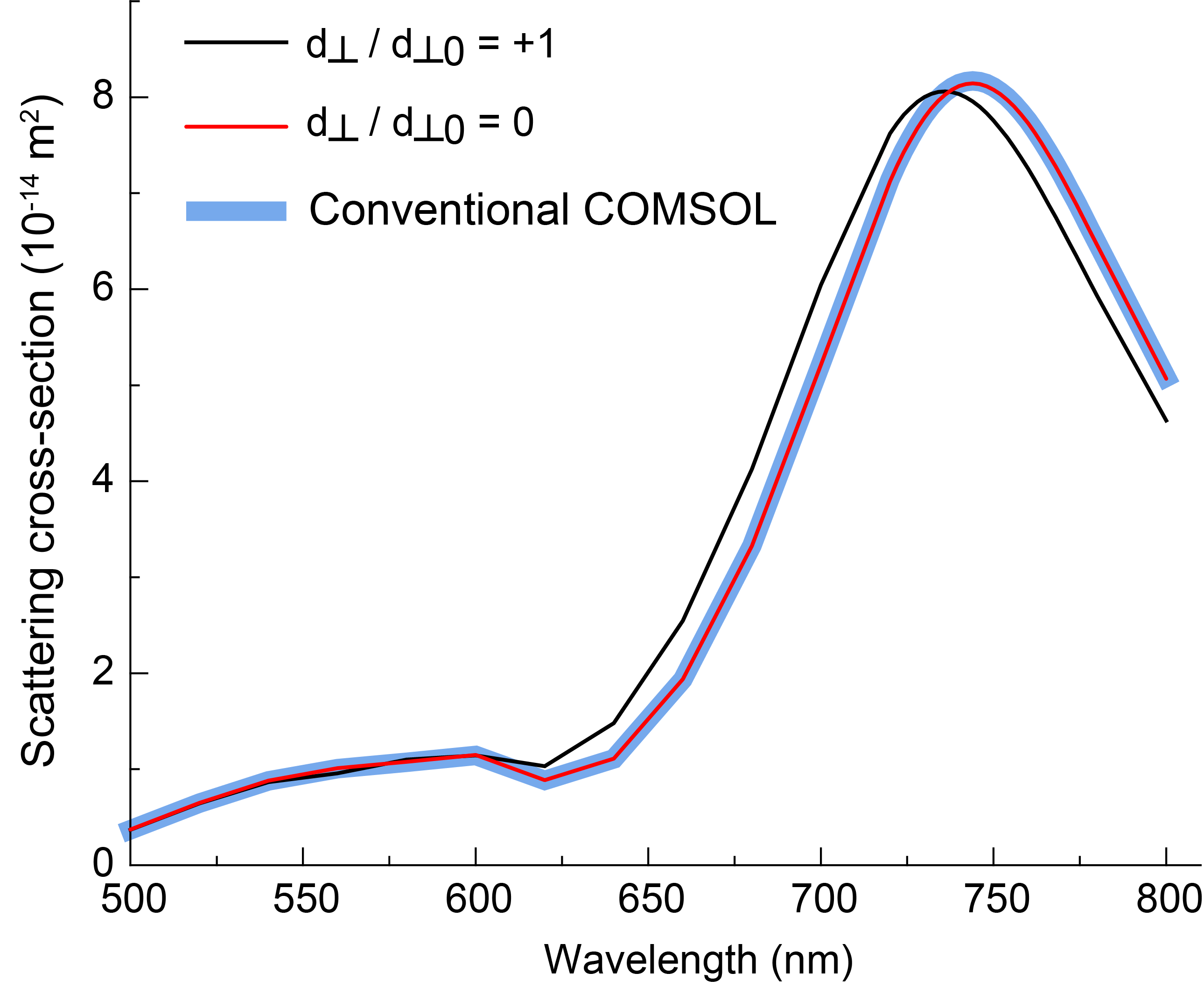}
    \caption{Scattering cross-section spectra of a single NPoM with $h=2$~nm. The results of the non-local calculations with negligible non-locality $d_{\rm \perp}=0$ (red line) agree well with those of the conventional (local) calculations (thick blue line). The results obtained in a non-local model with $d_{\rm \perp}=d_{\rm \perp0}$ are clearly different (black line).}
    \label{sfig:veri}
\end{figure}

\pagebreak
\subsection{Second-order correction to the linear shift of the resonance}

\begin{figure}[h]
    \centering
    \includegraphics[width=0.48\linewidth]{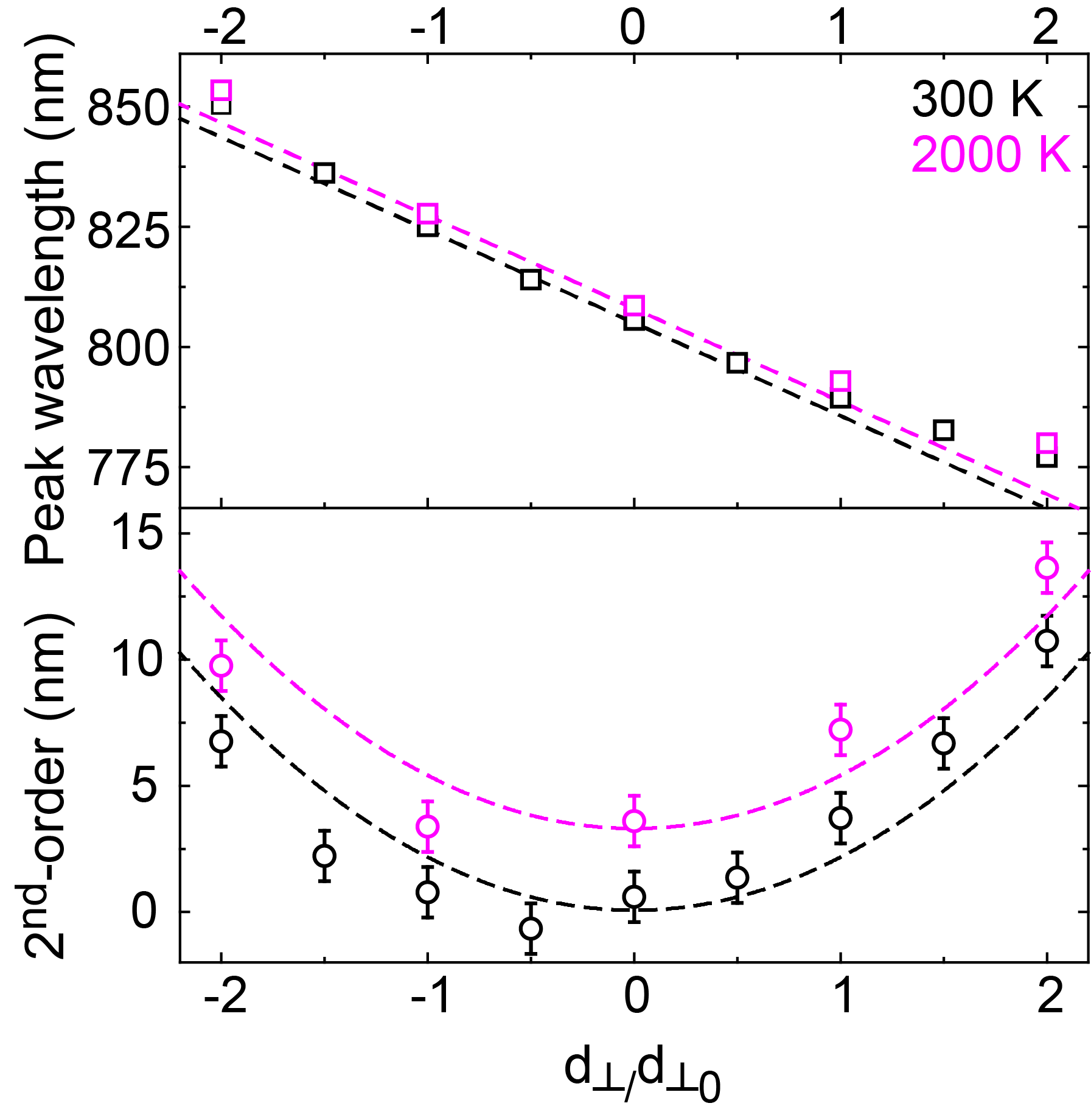}
    \caption{Detailed analysis of the role of non-locality in NPoM systems with cold (300~K) and laser-heated (2000~K) mirrors. Top panel: the dots show the peak positions extracted from the calculated spectra (cf. Fig.~4 in the main text), the parallel dashed lines are the linear fits with the shared slope. Bottom panel: the dots show the residuals (data from the top panel with linear fit subtracted) whereas the dashed lines illustrate the second-order correction to the linear approximation. 
    The gap width $h=1$~nm.}
    \label{sfig:2ndorder}
\end{figure}

\pagebreak
\subsection{Non-equilibrium NPoM systems: symmetry role}

The built-in asymmetry between the mirror and the particle in NPoM reportedly results in interesting emission properties \cite{Li2021}. In our approach, the asymmetry is further emphasized by the hot electron injection into the mirror and thus its elevated $T_e$. To verify the anticipated degree of asymmetry, we flip the electron temperatures and numerically compare a system depicted in Fig.~2 to its opposite, that is, where $T_e = 2000$~K on the particle while the mirror remains cold. As a measure for the asymmetry, we compare it with the unexcited NPoM where both the mirror and the particle are kept at $T_e = 300$~K. The results (Fig.~S5) indicate that the symmetric contribution to scattering dominates although the resonance shift is slightly stronger when the mirror is hot. This effect should vanish for larger nanoparticles, approaching the symmetric waveguide geometry whereas for even smaller ones a more pronounced asymmetric response can be expected.

\begin{figure}[h]
    \centering
    \includegraphics[width=0.48\linewidth]{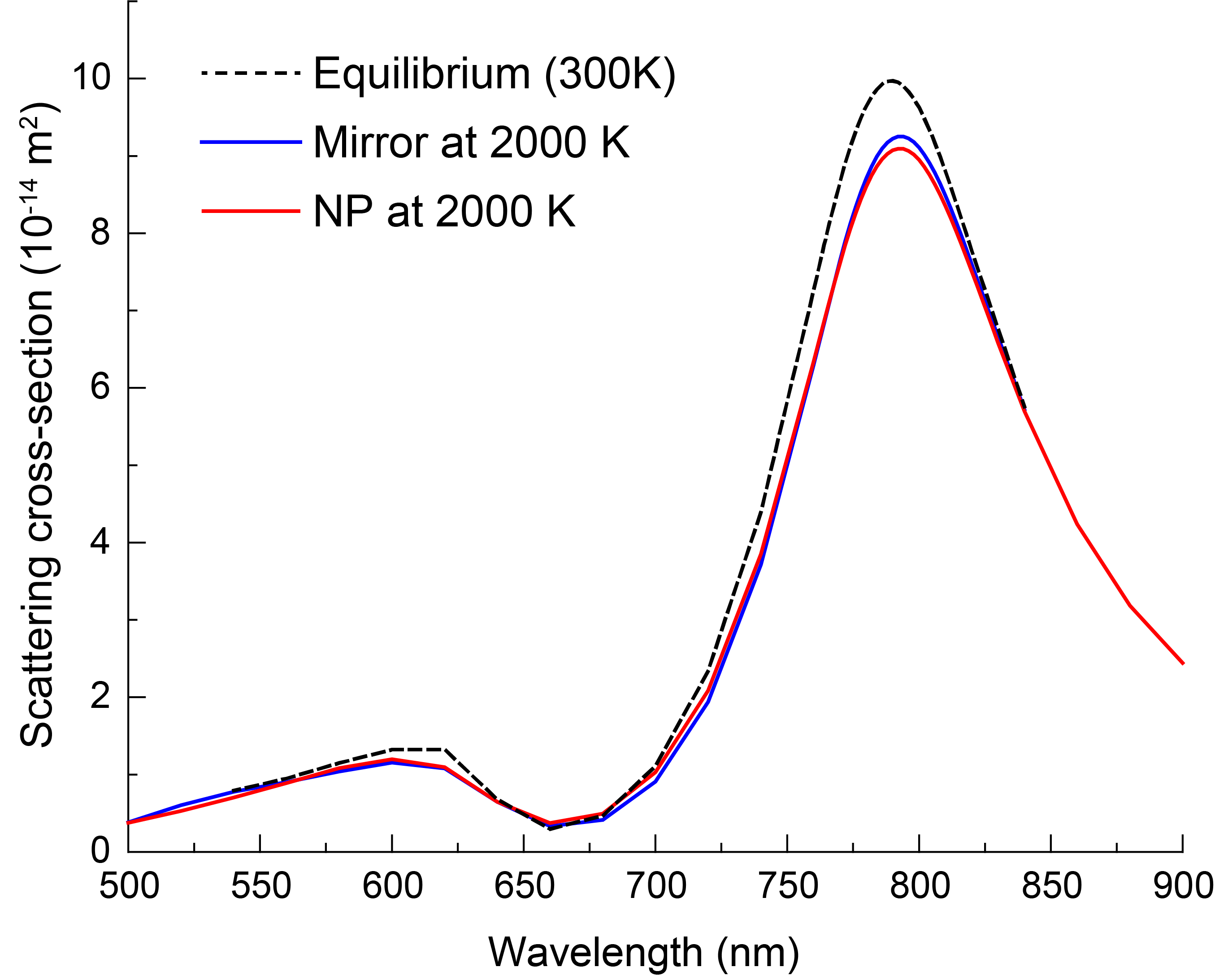}
    \caption{Calculated scattering cross-section spectra of a NPoM with $h=1$~nm when both nanoparticle and mirror are in the equilibrium at 300 K (black dashed line), as well as when the nanoparticle (red) or the mirror (blue) is laser-heated up to 2000~K. 
    }
    \label{sfig:asym}
\end{figure}

\pagebreak
\subsection{Dielectric function of Au at elevated temperatures}

\begin{figure}[h]
    \centering
    \includegraphics[width=0.7\linewidth]{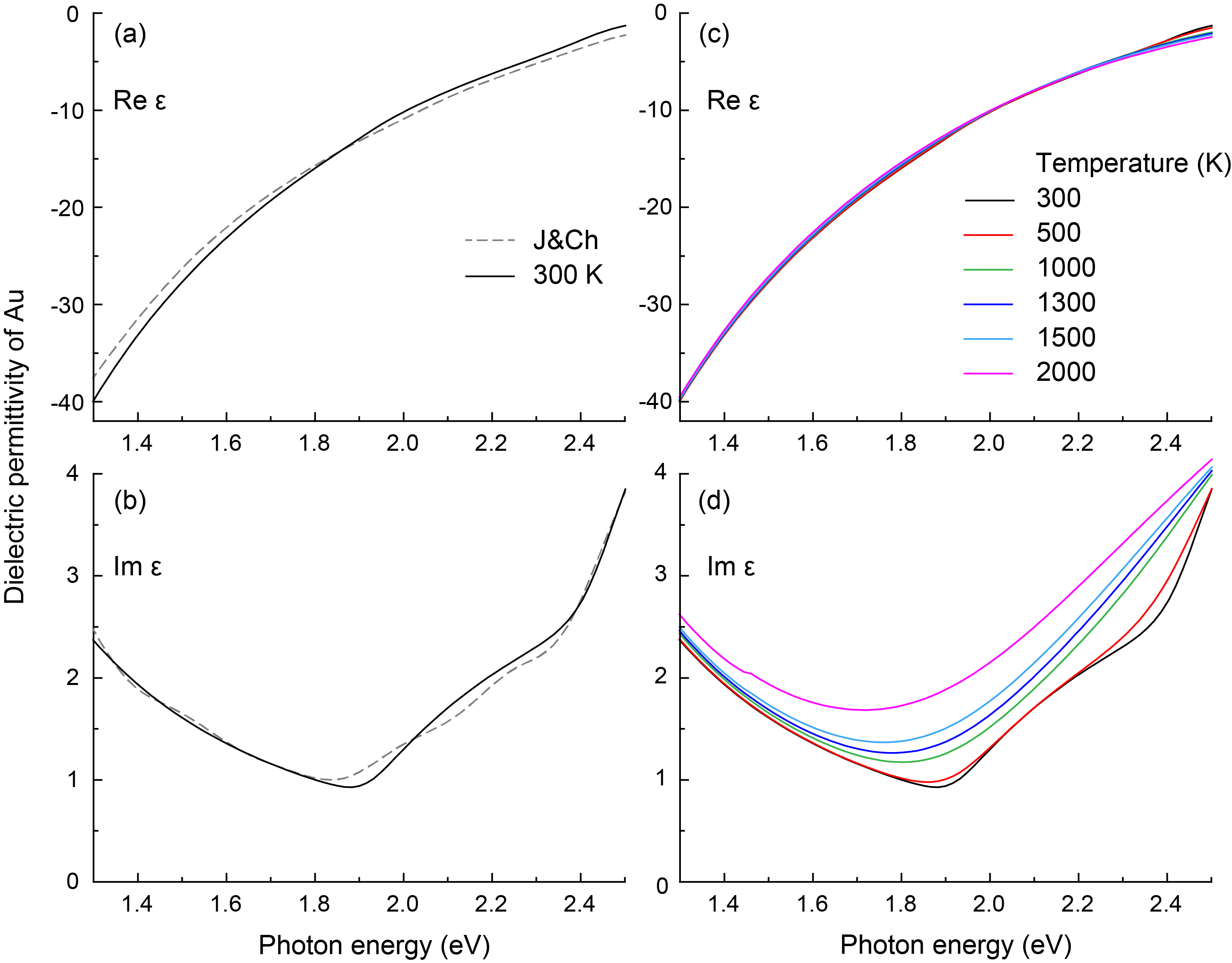}
    \caption{Calculated dielectric function of Au using the approach from Ref.~\cite{PhysRevB.12.557}. In panels (a-b), the results of our calculations for the real (a) and imaginary (b) parts are compared against the data from Johnson and Christy~\cite{JohnsonChristy1972} (dashed lines). In panels (c-d), the evolution of the calculated dielectric function with temperature is shown.}
    \label{sfig:epsilon}
\end{figure}

\pagebreak
\subsection{Linewidths and Q-factors in NPoM systems with non-locality}

\begin{figure}[h]
    \centering
    \includegraphics[width=0.7\linewidth]{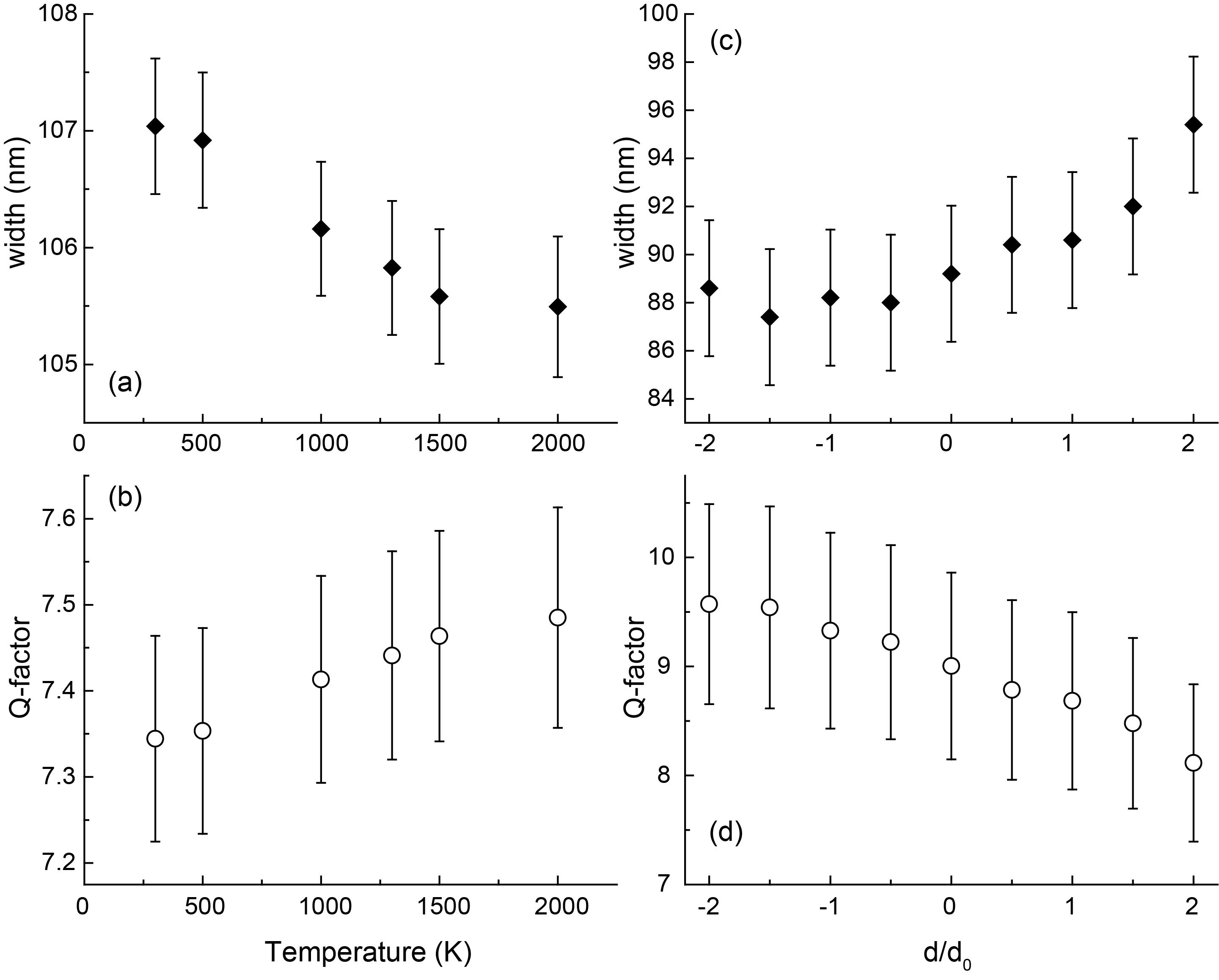}
    \caption{Linewidths (a,c) and recalculated Q-factors of the plasmon gap mode in non-local non-equilibrium NPoM systems. The linewidths are obtained from fitting the calculated scattering cross-section spectra with a Lorentzian in the vicinity of the resonance. The Q-factors are calculated as a ratio of the resonance frequency to the linewidth in the frequency domain, $Q=\omega_0/\Delta\omega$. As in the main text, in (a-b) the horizontal axis refers to the electronic temperature in the mirror whereas the nanoparticle remains at room temperature. In (c-d), the Feibelman non-locality $d_\perp$ is varied with respect to the $d_0$ values calculated in Ref.~\cite{RodriguezEcharri2021}.}
    \label{sfig:width}
\end{figure}

\newpage

\bibliography{aldito}

\end{document}